\documentclass[authorversion,pdftex, algorithms, LaTeX, 10pt, a4paper]{mdpi} 

\firstpage{1} 
\copyrightyear{2023}
\pdfoutput=1 

\usepackage[utf8]{inputenc}
\usepackage[autostyle=true]{csquotes} 
\usepackage[english]{babel}
\usepackage{graphicx}
\usepackage{mathtools}
\usepackage[ruled]{algorithm2e}
\usepackage{xcolor}
\usepackage{amsmath}
\usepackage{amsthm}
\usepackage{amssymb}
\usepackage{float}

\newcommand{\remove}[1]{}
\newcommand{\cP}{\mathcal{P}}
\newcommand{\cA}{\mathcal{A}}
\newcommand{\cX}{\mathcal{X}}
\newcommand{\cY}{\mathcal{Y}}
\newcommand{\oq}{\overline{q}}

\allowdisplaybreaks

\Title{Hardness and Approximability of Dimension Reduction on the Probability Simplex}

\Author{{Roberto Bruno} 
 \orcidA{}}

\AuthorNames{Roberto Bruno}

\address[1]{%
Department of Computer Science, University of Salerno, 84084 Fisciano, {Italy;} 
 rbruno@unisa.it}

\abstract{{Dimension reduction is a technique used to transform data from a high-dimensional space into a lower-dimensional space, aiming to retain as much of the original information as possible. This approach is crucial in many disciplines like engineering, biology, astronomy, and economics. In this paper, we consider the following dimensionality reduction instance}: 
Given an $n$-dimensional probability distribution $p$ and an integer $m<n$, 
we {aim to find} the 
$m$-dimensional probability distribution $q$ that is the closest to $p$, {using the Kullback--Leibler divergence as the measure of closeness}.
We prove that the problem is strongly NP-hard, and we {present}
an approximation algorithm for it.}


\begin{document}

\section{Introduction}\label{section:1}
\textit{{Dimension reduction}} \cite{B, SVM14} is a methodology {for mapping} data from a high-dimensional space to a lower-dimensional space, {while approximately preserving the original information content}. 
{This process is essential in fields such as engineering, biology, astronomy, and economics, where large datasets with high-dimensional points are common}.

It is often the case that the  
computational complexity of the algorithms employed to  
extract relevant 
information from these datasets  depends on the dimension
of the space where the points lie.
Therefore, it is important to find  a representation of the   data 
in a lower-dimensional space that still (approximately) preserves the 
information content  of the original data, as per 
given criteria. 

A special case of the general issue illustrated before arises when the 
elements of the dataset  are
$n$-dimensional probability distributions, and the problem is to approximate them
by lower-dimensional ones. This question has been 
extensively studied in different contexts.
{In \cite{AKMV15, Ca+}, the authors address the problem of dimensionality reduction on sets of probability distributions with the aim of preserving specific properties, such as pairwise distances. In \cite{Go}, Gokhale considers the problem of finding the
distribution that minimizes, subject to a set of linear constraints on
the probabilities, the ``discrimination information'' with respect to a
given probability distribution.
Similarly, in \cite{GT}, Globerson {et al.} address the dimensionality reduction problem by introducing a nonlinear method aimed at minimizing the loss of mutual information from the original data. In \cite{Le}, Lewis explores dimensionality reduction for reducing storage requirements and proposes an approximation method based on the maximum entropy criterion. Likewise, in \cite{ad}, Adler {et al.} apply dimensionality reduction to storage applications, focusing on the efficient representation of large-alphabet probability distributions.
More closely related to the dimensionality reduction that we deal with in this paper are the works \cite{COHEN, CLW, V12,cicalese_short_vectors}. In \cite{COHEN, CLW}, the authors address task scheduling problems where the objective is to allocate tasks of a project in a way that maximizes the likelihood of completing the project by the deadline. They formalize the problem in terms of random variables approximation by using the Kolmogorov distance as a measure of distance and present an optimal algorithm for the problem.  In contrast, in \cite{V12}, Vidyasagar defines a metric distance between
probability distributions on two distinct finite sets of possibly different cardinalities based on the Minimum Entropy Coupling (MEC) problem. Informally, in the MEC, given two probability distributions $p$ and $q$, one seeks to find a joint distribution $\phi$ that has $p$ and $q$ as marginal distributions and also has minimum entropy. Unfortunately, computing the MEC is NP-hard, as shown in \cite{KOVACEVIC2015369}. However, numerous works in the literature present efficient algorithms for computing couplings with entropy within a constant number of bits from the optimal value \cite{cicalese2019minimum, compton2022tighter, compton2023minimum, li2021efficient, sokota2024computing}.
We note that computing the coupling of a pair of distributions can be seen as essentially the inverse of dimension reduction. Specifically, given two distributions $p$ and $q$, one constructs a third, larger distribution $\phi$, such that $p$ and $q$ are derived from $\phi$ or, more formally, aggregations of $\phi$. In contrast, the dimension reduction problem addressed in this paper involves starting with a distribution $p$ and creating another, smaller distribution that is derived from $p$ or, more formally, is an aggregation of $p$.

Moreover, in \cite{V12}, the author demonstrates that, according to the defined metric, any optimal reduced-order approximation must be an aggregation of the original distribution. Consequently, the author provides an approximation algorithm based on the total variation distance, using an approach similar to the one we will employ in Section \ref{Approximation}. 
Similarly, in \cite{cicalese_short_vectors}, Cicalese {et al.} examine dimensionality reduction using the same distance metric introduced in \cite{V12}. They propose a general criterion for approximating $p$ with a shorter vector $q$, based on concepts from Majorization theory, and provide an approximation approach to solve the problem.

}
We also mention that analogous 
problems arise in  
\textit{{scenario reduction}} \cite{Ru}, where the problem is to (best)
approximate a given discrete distribution with another  
distribution with fewer atoms in compressing 
probability distributions \cite{Ga} and elsewhere \cite{CLW2,PU, PP}. Moreover, we recommend the following survey for further application examples \cite{Me}.

In this paper, we study the following instantiation of the general 
problem described before:
Given an $n$-dimensional probability distribution $p=(p_1, \ldots , p_n)$, and $m<n$,
find the $m$-dimensional probability distribution $q=(q_1, \ldots , q_m)$ that is the 
\textit{closest} to $p$, where the measure of closeness is the well-known 
relative entropy \cite{KL} (also known as Kullback--Leibler divergence). 
In Section \ref{sec:2}, we formally state the problem. In Section \ref{Hardness}, we prove that the problem 
is strongly NP-hard, and in Section \ref{Approximation},  we provide
an approximation algorithm  returning a solution whose distance from $p$ is at most 
1 plus the minimum possible distance.

\section{Statement of the Problem and Mathematical Preliminaries}\label{sec:2}

Let 
\begin{equation}
        \cP_n = \{p=(p_1,\dots,p_n)\mid p_1\geq \dots \geq p_n>0,   \sum_{i=1}^n p_i = 1\}
\end{equation}
be the $(n-1)$-dimensional probability simplex.
Given two probability distributions $p \in \cP_n$ and $q \in \cP_m$, with $m < n$, we say that $q$ is an \textit{{aggregation}} of $p$
if each component of $q$ can be expressed as the sum of distinct components of $p$. 
More formally, $q$ is an {aggregation} of $p$ if there exists a \emph{partition}
$\Pi=(\Pi_1, \ldots , \Pi_m)$ of $\{1, \ldots , n\}$ such that 
$q_i=\sum_{j\in \Pi_i}p_j$, for each $i=1, \ldots , m$.
Notice that the aggregation operation corresponds to the following operation
on random variables: Given a random variable $X$ that takes value 
in a finite
set $\cX=\{x_1, \ldots , x_n\}$, such that $\Pr\{X=x_i\}=p_i$ for $i=1, \ldots , n$,
\textit{any}  
function $f:\cX\mapsto \cY$, with $\cY=\{y_1,\dots,y_m\}$ and $m<n$, induces a random variable
$f(X)$ whose probability distribution $q=(q_1, \ldots , q_m)$ is 
an aggregation of $p$. Dimension reduction in random variables
through the application of
deterministic functions is a common technique in the area (e.g., 
\cite{La,V12,COHEN}).
Additionally, the problem arises also in the area of
\lq\lq hard clustering''~\cite{KMN} where one seeks
 a deterministic mapping $f$ 
from  data, generated by an r.v. $X$ taking values in a set $\cX$, 
to ``labels'' in some set $\cY$, where typically 
$|\cY|\ll |\cX|$. 

\smallskip
For any probability distribution $p \in \cP_n$ and an integer $m<n$, let us denote by $\cA_m(p)$
the set of all $q \in \cP_m$ that are aggregations of $p$.
{Our goal is to solve the following optimization problem}:

\begin{Problem}\label{prob}
Given $p \in \cP_n$ and  $m<n$, find
$q^* \in \cA_m(p)$  
such that
\begin{equation}\label{argmin}
   \min_{q\in\cA_m(p)} D(q \Vert p)=D(q^*\Vert p), 
\end{equation}
where $D(q \Vert p)$ is 
the relative entropy \cite{KL}, given by
$$D(q \Vert p) =  \sum_{i=1}^m q_i\log{\frac{q_i}{p_i}},$$
and the logarithm is of base $2$.
\end{Problem}

An additional  motivation to study Problem \ref{prob} comes from the fundamental 
paper \cite{Shore}, in which the principle of minimum relative entropy 
(called therein \textit{{minimum cross entropy principle}}) is derived in an 
axiomatic manner.
The 
principle states that, of the distributions $q$ that satisfy given  
constraints (in our case, that $q\in\cA_m(p)$), one  should choose the one with the least 
relative entropy \lq\lq distance'' from the prior $p$. 

\smallskip
Before establishing the computational complexity of 
the Problem \ref{prob}, we present a simple lower bound on the optimal value.

\begin{Lemma}\label{lemma-lb}
For each $p\in \cP_n$ and $q\in \cP_m$, $m<n$,  it holds that
\begin{equation}
    D(q \Vert p) \geq D(lb(p) \Vert p) = -\log\left(\sum_{i=1}^m p_i\right).
\end{equation}
where
\begin{equation}\label{eq:lb}
    lb(p) = \left(\frac{p_1}{\sum_{i=1}^m p_i},\dots,\frac{p_m}{\sum_{i=1}^m p_i}\right)\in \cP_m.
\end{equation}
\end{Lemma}
\begin{proof}
Given an arbitrary $p\in \cP_n$, one can see that
$$D(lb(p) \Vert p) = -\log\left(\sum_{i=1}^m p_i\right).$$
Moreover, for \textit{any} $p\in \cP_n$ 
\textit{and} $q\in \cP_m$,  the Jensen inequality
applied to the $\log$ function gives the following:
\begin{equation*}
-D(q||p)=\sum_{i=1}^mq_i\log\frac{p_i}{q_i}
          \leq\log\left (\sum_{i=1}^mp_i\right ).
\end{equation*}
\end{proof}

\section{Hardness}\label{Hardness}
In this section, we prove that the
optimization problem  (\ref{argmin}) described in Section \ref{section:1} is strongly NP-hard.
{We accomplish this by reducing the problem} from the \textsc{3-Partition} problem, a well-known strongly NP-hard problem \cite{Gar}, {described as follows}.

\smallskip
\textsc{3-Partition}: Given a multiset $S = \{a_1,\dots,a_n\}$ of $n=3m$ positive integers for which $\sum_{i=1}^n a_i = mT$, for some $T$, 
the problem is to decide whether $S$ can be partitioned into $m$ triplets such that the sum of each triple is exactly $T$.
More formally,  the problem is to decide whether there exist 
$S_1,\dots,S_m\subseteq S$ such that the following conditions hold:

\begin{align*}
    \sum_{a \in S_j} a &= T, \quad \forall j \in \{1,\dots,m\},\\
    S_i \cap S_j &= \emptyset, \quad \forall i \neq j,\\
    \bigcup_{i=1}^m S_i &= S,\\
    |S_i| &= 3, \quad \forall i\in \{1,\dots,m\}.
\end{align*}

\begin{Theorem}
The \textsc{3-Partition}
problem can be reduced in polynomial time to the problem 
of finding the aggregation $q^*\in \cP_{m}$ of 
some  $p\in \cP_{n}$, 
for which 
$$D(q^* \Vert p)=\min_{q\in\cA_m(p)} D(q \Vert p).$$
\end{Theorem}

\begin{proof}
{The idea behind the following reduction can be summarized as follows: given an instance of \textsc{3-Partition}, we transform it into a probability distribution $p$ such that the lower bound $lb(p)$ is an aggregation of $p$ if and only if the original instance of \textsc{3-Partition} admits a solution.}
Let  an 
arbitrary instance of \textsc{3-Partition} be given, that is, let $S$ be a 
multiset $\{a_1,\dots,a_n\}$ of $n=3m$ positive integers with 
$\sum_{i=1}^n a_i = mT$. Without loss of generality, we assume that the integers $a_i$ are ordered in a non-increasing fashion. 
We construct a valid instance $p$ of our Problem \ref{prob} as follows. We 
set $p \in \cP_{n+m}$ as follows:
\begin{equation}\label{defp}
    p = \left( \underbrace{ \frac{1}{m+1},\dots,\frac{1}{m+1}}_{m \mbox{ \footnotesize times}},
    \frac{a_1+2T}{(m+1)7mT},\dots, \frac{a_n+2T}{(m+1)7mT}
    \right).
\end{equation}
{Note that} 
 $p$ is a probability distribution. In fact, since $n=3m$, we have 
$$
\sum_{i=1}^n \frac{a_i+2T}{(m+1)7mT} = 
\frac{1}{(m+1)7mT}\left(\sum_{i=1}^n (a_i +2T)\right)=\frac{7mT}{(m+1)7mT} = \frac{1}{m+1}.
$$
Moreover, from (\ref{eq:lb}) and (\ref{defp}), the probability distribution $lb(p)\in \cP_m$ associated to  $p$ is as follows:
\begin{equation}\label{eq:unif}
    lb(p) = \left(\frac{p_1}{\sum_{j=1}^m p_j},\dots,\frac{p_m}{\sum_{j=1}^m p_j}\right)=\left(\frac{1}{m},\dots,\frac{1}{m}\right).
\end{equation}

To prove the theorem, we show  that the starting  instance of 
\textsc{3-Partition} is a \textsc{Yes} instance  \textit{{if and only if}}
it holds that
\begin{equation}\label{iff}
\min_{q\in\cA_m(p)} D(q \Vert p)=\log \frac{m+1}{m},
\end{equation}
where $p$ is given in (\ref{defp}).

We begin  by assuming the given   instance of \textsc{3-Partition}
is a \textsc{Yes} instance, that is, 
there is a partition of $S$ into triplets $S_1,\dots,S_m$ such that 
\begin{equation}\label{eq:ipotesi_1}
    \sum_{a_i\in S_j} a_i = T, \quad \forall j \in \{1,\dots,m\},
\end{equation}
and we show that $\min_{q\in\cA_m(p)} D(q \Vert p)=\log \frac{m+1}{m}$.
By Lemma \ref{lemma-lb}, (\ref{defp}), and   equality (\ref{eq:unif}), we have
\begin{equation}\label{low}
\min_{q\in\cA_m(p)} D(q \Vert p) \geq D(lb(p)\Vert p)=\sum_{i=1}^m\frac{1}{m}\log\frac{1/m}{1/(m+1)}=
\log \frac{m+1}{m}.
\end{equation}
From (\ref{eq:ipotesi_1}), we have 
\begin{align}
    \sum_{a_i \in S_j} \frac{a_i+2T}{(m+1)7mT} 
    &= \frac{T}{(m+1)7mT} + \sum_{a_i \in S_j} \frac{2T}{(m+1)7mT} \nonumber\\
    &= \frac{T}{(m+1)7mT} +  \frac{6T}{(m+1)7mT}\nonumber \\ 
    &=\frac{1}{(m+1)m}, \qquad \forall j \in \{1,\dots,m\}. \label{sumai}
\end{align}
Let us define $q'\in\cP_m$ as follows:

\begin{equation}
    q' = \left(\frac{1}{m+1} + \sum_{a_i \in S_1} \frac{a_i+2T}{(m+1)7mT},\dots,\frac{1}{m+1} + \sum_{a_i \in S_m} \frac{a_i+2T}{(m+1)7mT}\right),
\end{equation}
where, by (\ref{sumai}), 
\begin{equation}\label{qisaggre}
\sum_{a_i \in S_j} \frac{a_i+2T}{(m+1)7mT} = \frac{1}{(m+1)m}, \quad \forall j \in \{1,\dots,m\}.
\end{equation}
From (\ref{qisaggre}) and from the fact that $S_1, \ldots, S_m$ are a partition
of $\{a_1, \ldots , a_n\}$, we obtain  $q'\in \cA_m(p)$, that is, 
$q'$ is a valid  aggregation of $p$ (cfr., (\ref{defp})). Moreover, 
 
 $$q' = \left(\frac{1}{m},\dots,\frac{1}{m}\right),$$
 and $D(q'\Vert p)=\log \frac{m+1}{m}$.
 Therefore, by (\ref{low}) and that $q'\in \cA_m(p)$, we obtain
 $$\min_{q\in\cA_m(p)} D(q \Vert p)=\log \frac{m+1}{m},$$
 as required.


\smallskip
To prove the opposite implication, 
we assume that $p$ (as given in (\ref{defp}))  is a \textsc{Yes} instance, that is,
\begin{equation}\label{inversa}
\min_{q\in\cA_m(p)} D(q \Vert p)=\log \frac{m+1}{m}.
\end{equation}

We show that the original instance of \textsc{3-Partition} is also a \textsc{Yes} instance, that is, there is a partition of $S$ into triplets $S_1,\dots,S_m$ such that 
\begin{equation}
    \sum_{a_i\in S_j} a_i = T, \quad \forall j \in \{1,\dots,m\}.
\end{equation}
Let $q^*$ be the element in $\cA_m(p)$ that achieves the minimum in (\ref{inversa}). Consequently,
we have 
\begin{align}
\log \frac{m+1}{m}&=D(q^*\Vert p)= \sum_{i=1}^mq^*_i\log\frac{q^*_i}{p_i}
=\sum_{i=1}^mq^*_i\log\frac{1}{p_i}- H(q^*)\nonumber \\
&=\log(m+1)- H(q^*) \qquad (\mbox{from (\ref{defp}))}, \label{cons}
\end{align}
where $H(q^*)=-\sum_{i=1}^m q^*_i\log q^*_i$ is the Shannon entropy of $q^*$. 
From (\ref{cons}), we obtain that $H(q^*)=\log m$; hence, $q^*=(1/m, \ldots , 1/m)$
(see \cite{Cov},  Thm. 2.6.4).
Recalling that $q^*\in \cA_m(p)$, we obtain that
the uniform distribution 
\begin{equation}\label{eq:uniform}
    \left(\frac{1}{m},\dots,\frac{1}{m}\right)
\end{equation}
is an \textit{aggregation} of $p$. We note that the first $m$ components of $p$, as 
defined in (\ref{defp}), cannot be aggregated among them to obtain (\ref{eq:uniform}), because $2/(m+1) > 1/m$, for $m>2$. Therefore, in order to obtain (\ref{eq:uniform}) as an aggregation of $p$, there must exist a partition $S_1,\dots, S_m$ of $S=\{a_1, \ldots , a_n\}$ for which 
\begin{equation}\label{eq:ipotesi_2}
    \frac{1}{m+1} + \sum_{a_i \in S_j} \frac{a_i+2T}{(m+1)7mT} = \frac{1}{m}, \quad \forall j \in \{1,\dots,m\}.
\end{equation}
From (\ref{eq:ipotesi_2}), we obtain
\begin{equation}
    \sum_{a_i \in S_j} \frac{a_i+2T}{(m+1)7mT} = \frac{1}{m(m+1)},\quad \forall j \in \{1,\dots,m\}.
\end{equation}
From this, it follows that 
\begin{equation}\label{eq:consequenza}
     2T|S_j| + \sum_{a_i \in S_j} a_i = 7T,\quad \forall j \in \{1,\dots,m\}.
\end{equation}
We note that,  for (\ref{eq:consequenza}) to be true, there cannot exist any $S_j$ for which $|S_j| \neq 3$. Indeed, if there were a subset $S_j$ 
for which $|S_j|\neq 3$, there would be at least a subset $S_k$ for which $|S_k|>3$. 
Thus, for such an  $S_k$, we would  have
\begin{equation*}
    2T|S_k| + \sum_{a_i \in S_k} a_i \geq 8T +\sum_{a_i \in S_k} a_i > 7T,
\end{equation*}
contradicting (\ref{eq:consequenza}).
Therefore, it holds that
\begin{equation}\label{eq:triplets}
   |S_j| = 3,\quad\forall j\in\{1,\dots,m\}. 
\end{equation}
Moreover, from (\ref{eq:consequenza}) and (\ref{eq:triplets}), we obtain 
\begin{equation}\label{eq:sum}
    \sum_{a_i \in S_j} a_i = 7T - 2T|S_j| = T,\quad\forall j \in\{1,\dots,m\}.
\end{equation}
Thus, from (\ref{eq:sum}), it follows that
the subsets $S_1, \ldots , S_m$ give a partition of $S$ into triplets,
such that 
\begin{equation*}
    \sum_{a_i\in S_j} a_i = T, \quad \forall j \in \{1,\dots,m\}.
\end{equation*}
Therefore, the starting 
instance of \textsc{3-Partition} is a \textsc{Yes} instance. 
\end{proof}

\section{Approximation}\label{Approximation}
Given $p\in \cP_n$ and  $m<n$,  let $OPT$ denote the optimal value of the optimization 
problem (\ref{argmin}), that is
\begin{equation}\label{opt}
OPT=\min_{q\in\cA_m(p)} D(q \Vert p).
\end{equation}
In this section,  we design a greedy algorithm to compute an aggregation 
$\oq\in \cA_m(p)$ of $p$ such that 
\begin{equation}\label{approx}
D(\oq \Vert p)<OPT+1.
\end{equation}

The idea behind our  algorithm is to
see the problem of computing 
an  aggregation $q\in \cA_m(p)$ as 
a bin packing problem with ``overstuffing'' (see \cite{DOP} and
references therein quoted), {which is a bin packing where overfilling of bins is possible}. 
In  the classical bin packing problem, 
one is given a set of items, with their associated weights, and a set of bins with their associated capacities (usually, equal for all bins). The objective is to place all the items in the bins, trying to minimize a given  
cost function.

In our case, we have $n$ items (corresponding to the components of  $p$) with weights $p_1,\dots,p_n$, respectively,  and $m$ bins, corresponding to the components of  $lb(p)$ (as defined in (\ref{eq:lb})) with capacities $lb(p)_1,\dots, lb(p)_m$. Our objective is to place all the $n$ components of $p$ into the $m$ bins \textit{without exceeding} the capacity $lb(p)_j$ of each bin  $j$,
$j=1, \ldots , m$,  by more than $(\sum_{i=1}^m p_i)lb(p)_j$. {For such a purpose, the idea behind Algorithm \ref{alg} is quite straightforward. It behaves like a classical First-Fit bin packing: to place the $i^{th}$ item, it chooses the first bin $j$ in which the item can be inserted without exceeding its capacity by more than $(\sum_{i=1}^m p_i)lb(p)_j$. 
	 In the following, we will show that such a bin always exists and that} fulfilling this objective is sufficient to ensure the
approximation guarantee (\ref{approx}) we are seeking.

\vspace{6pt}
\begin{algorithm}[H]
{
\label{alg}
\caption{{GreedyApprox} 
}
    1. Compute   $lb(p) = (p_1/\sum_{j=1}^m p_j,\dots,p_m/\sum_{j=1}^m p_j)$;\\
    2. Let  $lb_j^i$ be the content of bin $j$ after the first $i$ components of $p$ have been placed 
    ($lb_j^0 = 0$  for each 
    $j \in \{1,\dots,m\}$);\\
    3. For $i = 0, \dots, n-1$\\
    \hspace{1cm} Let $j$ be the smallest bin index for which holds that \\
    \hspace{1cm} $lb_j^i + p_{i+1} < (1 + \sum_{j=1}^m p_j)lb(p)_j$,  place $p_{i+1}$
    into the $j$-th bin:\\
    \hspace{1.5cm} 
    $lb_j^{i+1} = lb_j^i + p_{i+1},$\\
    \hspace{1.5cm} 
    $lb_k^{i+1} = lb_k^i$, for each $k \neq j$;\\
    4. Output  $\oq=(lb_1^n, \ldots , lb_m^n)$.
 }
\end{algorithm}
\vspace{0.5cm}
The step 3  of {{GreedyApprox}} 
operates as in the classical  First-Fit bin packing algorithm.
Therefore, 
it can be implemented to run in $O(n\log m)$ time, as discussed in \cite{hochba1997}.  {In fact, each iteration of the loop in step 3 can be implemented in $O(\log m)$-time by using a balanced binary search tree with height $O(\log m)$ that has a leaf for each bin and in which each node keeps track of the largest remaining capacity of all the bins in its subtree.}

\begin{Lemma}
\label{th:soundness} 
{{GreedyApprox}} computes  a valid aggregation $\oq\in\cA_m(p)$ of $p\in \cP_n$. 
Moreover, it holds that
\begin{equation}\label{boundalb}
    D(\oq \Vert lb(p)) < \log\left(1 + \sum_{j=1}^m p_j\right).
\end{equation}
\end{Lemma}

\begin{proof}
We first prove that each component $p_i$ of $p$ is placed in some bin. This implies that $\oq\in \cA_m(p)$.

For each step $i=0,\dots,m-1$, there is always a bin in which 
the algorithm places $p_{i+1}$.
In fact,  the capacity $lb(p)_j$ of bin $j$ satisfies the 
relation:
$$
lb(p)_j= \frac{p_j}{\sum_{\ell=1}^m p_{\ell}} > p_j, \quad \forall j \in \{1,\dots,m\}.
$$
Let us consider an arbitrary step $m\leq i < n$, 
in which the algorithm has placed the first $i$ components of $p$ 
and  needs to place $p_{i+1}$ into some bin. 
We show that, in this case also, there is always a bin $j$ in which 
the algorithm places the item $p_{i+1}$,
\textit{without exceeding} the capacity 
$lb(p)_j$ of the bin $j$ by more than $(\sum_{\ell=1}^m p_{\ell})lb(p)_j$. 

First, notice that in each step $i$,  $m\leq i < n$,
there is at least a bin $k$ whose content $lb_k^i$ \textit{does not} exceed its capacity $lb(p)_k$; that is, 
for which $lb_k^i < lb(p)_k$ holds.
Were this the opposite, 
 for all bins $j$, we would have  $lb_j^i \geq lb(p)_j$; then, we would also have 
\begin{equation}\label{eq:absurd}
    \sum_{j=1}^m lb_j^i \geq \sum_{j=1}^m lb(p)_j = 1.
\end{equation}
However, this is not possible since we have placed only the first $i<n$ components of $p$, and therefore,  it holds that
\begin{equation*}
    \sum_{j=1}^m lb_j^i = \sum_{j=1}^i p_j < \sum_{j=1}^n p_j = 1,
\end{equation*}
contradicting (\ref{eq:absurd}).
Consequently, let $k$ be the smallest integer for which the 
content of the $k$-th bin
does not exceed its capacity, i.e., for which 
$lb_k^i < lb(p)_k$. For such a bin $k$, we obtain
\begin{align}\label{eq:bin_k}
    \left(1+\sum_{j=1}^m p_j\right)lb(p)_k &= lb(p)_k + \left(\sum_{j=1}^m p_j\right)lb(p)_k\nonumber\\
    &= lb(p)_k + \left(\sum_{j=1}^m p_j\right) \frac{p_k}{\sum_{j=1}^m p_j}\nonumber\\
    &= lb(p)_k + p_k\nonumber\\
    &> lb_k^i + p_k\hspace{1cm}\mbox{(since $ lb(p)_k>lb_k^i$)}\nonumber\\
    &\geq lb_k^i + p_{i+1} \hspace{1cm}\mbox{(since $p_k \geq p_{i+1}$)}.
\end{align}
Thus, from (\ref{eq:bin_k}), one derives 
that the algorithm   places $p_{i+1}$ into the bin $k$ without exceeding its capacity $lb(p)_k$ by more than $(\sum_{j=1}^m p_j)lb(p)_k$.

{The reasoning applies to each $i<n$, thus proving that GreedyApprox correctly assigns each component $p_i$ of $p$ to a bin, effectively computing an aggregation of $p$}. Moreover, from the instructions 
of step 3 of {GreedyApprox}, 
the output is  an aggregation $\oq=(\oq_1, \ldots , \oq_m)\in \cA_m(p)$, 
for which the following crucial relation holds:
\begin{equation}\label{eq:invariante}
    \oq_i < \left(1 + \sum_{j=1}^m p_j\right)lb(p)_i, \quad\forall i \in \{1,\dots,m\}.
\end{equation}

\smallskip
Let us now prove that $ D(\oq \Vert lb(p)) < \log\left(1 + \sum_{j=1}^m p_j\right)$.
We have
\begin{align*}
    D(\oq \Vert lb(p)) &= \sum_{i=1}^m \oq_i \log\frac{\oq_i}{lb(p)_i}\\
    &<  \sum_{i=1}^m \oq_i \log \frac{(1 + \sum_{j=1}^m p_j)lb(p)_i}{lb(p)_i} \hspace{1cm}\mbox{(from (\ref{eq:invariante}))}\\
    &= \log\left(1 + \sum_{j=1}^m p_j\right).
\end{align*}
\end{proof}
We need the following technical lemma to show the approximation guarantee of {GreedyApprox}.
\begin{Lemma} \label{lemma:rel_entropy_sum}
Let $q \in \cP_m$ and $p \in \cP_n$ be two arbitrary probability distributions with $m < n$. It holds that 
\begin{equation}
    D(q \Vert p) = D( q \Vert lb(p)) + D(lb(p) \Vert p),
\end{equation}
where $lb(p) =(lb(p)_1, \ldots , lb(b)_m)= (p_1/\sum_{i=1}^m p_i, \dots, p_m /\sum_{i=1}^m p_i)$.
\end{Lemma}

\begin{proof}
\begin{align*}
    D(q \Vert p) &= \sum_{i=1}^m q_i\,\log \frac{q_i}{p_i}
    = \sum_{i=1}^m q_i\,\log \frac{q_i}{p_i\,\frac{\sum_{j=1}^m p_j}{\sum_{j=1}^m p_j}}\\
    &= \sum_{i=1}^m q_i\,\log \frac{q_i}{\frac{p_i}{\sum_{j=1}^m p_j}} + \sum_{i=1}^m q_i\,\log \frac{1}{\sum_{j=1}^m p_j}\\
    &= \sum_{i=1}^m q_i\,\log \frac{q_i}{lb(p)_i} + \sum_{i=1}^m q_i\,\log\frac{1}{\sum_{j=1}^m p_j}\hspace{1cm}\mbox{(since $lb(p)_i = p_i / \sum_{j=1}^m p_j$)}\\
    &= D(q \Vert lb(p)) + \log \frac{1}{\sum_{j=1}^m p_j}=  D(q \Vert lb(p)) +  D(lb(p) \Vert p).
\end{align*}
\end{proof}
The following theorem is the main result of this section.
\begin{Theorem}\label{th:approximation}
For any $p\in \cP_n$ and $m<n$,  {{GreedyApprox}}  produces an aggregation 
$\oq\in \cA_m(p)$ of $p$ such that
\begin{equation}
D(\oq\Vert p)<OPT +1,
    \end{equation}
where
$OPT=\min_{q\in\cA_m(p)} D(q \Vert p).$
\end{Theorem}

\begin{proof}
From Lemma \ref{lemma:rel_entropy_sum}, we have
\begin{equation}
    D(\oq \Vert p) = D(\oq \Vert lb(p)) + D(lb(p) \Vert p),
\end{equation}
and from Theorem \ref{th:soundness}, we know that the produced aggregation $\oq$  of $p$ satisfies the relation
\begin{equation}\label{eq:theorem_result}
    D(\oq \Vert lb(p)) < \log \left(1 + \sum_{j=1}^m p_j\right).
\end{equation}
Putting it all together, we obtain:
\begin{align*}
     D(\oq \Vert p) &= D(\oq \Vert lb(p)) + D(lb(p) \Vert p)\\
     &<\log \left(1 + \sum_{j=1}^m p_j\right) + D(lb(p) \Vert p) \hspace{1cm}\mbox{(from (\ref{eq:theorem_result}))}\\
     &= \log\left(1 + \sum_{j=1}^m p_j\right)-\log \left(\sum_{j=1}^m p_j\right)\\
     &<-\log \left(\sum_{j=1}^m p_j\right) + 1 \hspace{1cm}\mbox{(since $1 + \sum_{j=1}^m p_j < 2$)}\\
     &\leq OPT +1 \hspace{3.3cm}\mbox{(from Lemma \ref{lemma-lb}).}
\end{align*}
\end{proof}

\section{Concluding Remarks}
{In this paper, we examined the problem of approximating $n$-dimensional probability distributions with $m$-dimensional ones using the Kullback--Leibler divergence as the measure of closeness. We demonstrated that this problem is strongly NP-hard and introduced an approximation algorithm for solving the problem with guaranteed performance.}

Moreover, we conclude by pointing out that the analysis of {GreedyApprox} presented in Theorem \ref{th:approximation} is tight. 
Let $p \in \cP_3$ be 
\begin{equation*}
    p = \left(\frac{1}{2}-\epsilon, \frac{1}{2}-\epsilon, 2\epsilon\right),
\end{equation*}
where $\epsilon>0$. 
The application of {{GreedyApprox}} on $p$ produces  the 
aggregation $\oq \in \cP_2$ given by
\begin{equation*}
    \oq = (1-2\epsilon, 2\epsilon),
\end{equation*}
whereas one can see that the optimal aggregation $q^* \in \cP_2$ is equal to
\begin{equation*}
    q^* = \left(\frac{1}{2}+\epsilon, \frac{1}{2}-\epsilon\right).
\end{equation*}
Hence, for $\epsilon \to 0$, we have 
\begin{equation*}
    D(\oq \Vert p) = (1-2\epsilon)\log \frac{1-2\epsilon}{\frac{1}{2}-\epsilon} + 2\epsilon \log \frac{2\epsilon}{\frac{1}{2}-\epsilon} \to 1,
\end{equation*}
while 
\begin{equation*}
    OPT=D(q^* \Vert p) = \left(\frac{1}{2}+\epsilon\right)\log \frac{\frac{1}{2}+\epsilon}{\frac{1}{2}-\epsilon} + \left(\frac{1}{2}-\epsilon\right)\log \frac{\frac{1}{2}-\epsilon}{\frac{1}{2}-\epsilon} \to 0.
\end{equation*}
Therefore, to improve our approximation guarantee, 
one should use a bin packing heuristic
different from the First-Fit as employed in {{GreedyApprox}}. 
{{Another interesting open problem is to provide an approximation algorithm
with a (small) multiplicative approximation guarantee}. }
{However, both problems mentioned above 
 would probably require a different approach}, and we leave that to future investigations.

Another interesting line of research would be to extend our findings to 
different divergence measures (e.g., \cite{Sason} and references quoted therein).



\vspace{6pt}

\reftitle{References}

\end{document}